\documentclass[preprint,showpacs,12pt]{revtex4}
\usepackage{amsmath,amssymb,amsfonts}
\usepackage{graphicx}
\usepackage{hhline}
\usepackage{bm}
\usepackage{amsmath}
\usepackage{epsfig}

\begin{document}
\begin{flushleft}
CERN-PH-TH/2011-260\\
LYCEN-2011-13 \\
\end{flushleft}
\title{Neutrino quasielastic interaction and nuclear dynamics}
\author {M. Martini}
\affiliation{CEA/DAM/DIF, F-91297 Arpajon, France}
\author {M. Ericson}
\affiliation{Universit\'e de Lyon, Univ. Lyon 1,
 CNRS/IN2P3, IPN Lyon, F-69622 Villeurbanne Cedex, France}
\affiliation{Physics Department, Theory Unit, CERN, CH-1211 Geneva, Switzerland}
\author {G. Chanfray}
\affiliation{Universit\'e de Lyon, Univ. Lyon 1,
 CNRS/IN2P3, IPN Lyon, F-69622 Villeurbanne Cedex, France}

\begin{abstract}
 
We investigate the double differential neutrino-carbon quasielastic cross sections as measured by the MiniBooNE experiment. 
Our present treatment incorporates relativistic corrections in the nuclear response functions and includes the multinucleon component. 
We confirm our previous conclusion that it is possible to account for all the data without any modification of the axial mass. 
We also introduce the $Q^2$ distribution for charged and neutral current. 
The data point at a sizable multinucleon component beside the genuine quasielastic peak. 
They are also indicative of the  collective character of the nuclear response, of interest for hadronic physics.

\end{abstract}

\maketitle

\section{Introduction}
Recent data on neutrino-nucleus scattering have improved our understanding of the neutrino-nucleus interaction, 
needed for neutrino oscillation experiments where nuclear targets are involved.
Among the results on partial cross sections, the quasielastic one turns out to be very important \cite{AguilarArevalo:2010zc,AguilarArevalo:2010cx}. 
An outcome of these data was the display of an ``anomaly'' in the quasielastic cross section on $^{12}$C. This quantity can be fitted by a relativistic 
Fermi gas model only at the price of a modification of the axial form factor with an axial mass  $M_A=1.35$ GeV, instead of the usual value  $M_A=1.03$ GeV 
as measured in deuteron bubble chamber experiments. 
For nuclear physicists accustomed to the complexity of the many-body nuclear system, this anomaly is likely to reflect the many-body aspect of the problem. 
Indeed we have pointed out  \cite{Martini:2009uj,Martini:2010ex}  that, depending on the detection method,  certain types of inelastic events can simulate quasielastic ones. 
This is the case for interactions leading to a final state with two or more nucleons  ejected, 
if a quasielastic event is defined as one with only a muon in the final states, as in MiniBooNE. 
Multinucleon processes occur by nuclear correlations, with or without Delta excitation. We have argued  that this is the likely explanation of the anomaly 
showing that an evaluation can account for the excess cross section without any modification of the axial mass. After this suggestion, 
a number of articles \cite{Benhar:2010nx,Amaro:2010sd,Amaro:2011qb,Nieves:2011pp,Nieves:2011yp,Bodek:2011ps,Meucci:2011vd} 
have discussed the problem of multinucleon emission and whether it could account for the anomaly, with various conclusions, critical or supportive of our result.

Our previous works only dealt with the quasielastic cross section as a function of the neutrino energy. 
In the comparison with the experimental data, the uncertainties linked to the fact that the neutrino spectrum 
is broad are of experimental origin because the extraction of the energy dependence of the cross section involves a reconstruction of the neutrino energy whereas in the theoretical evaluation 
the neutrino energy is just an input.
In the present work we discuss the double differential cross section. 
This is a directly measured quantity, free from the uncertainty of  neutrino energy reconstruction. However there remains an uncertainty on the theoretical side 
because the measured double differential cross section refers to the broad  spectrum of neutrino energies. 
The theoretical predictions imply a convolution on this spectrum, which could be a source of error. Nevertheless a good agreement with  theory for the double 
differential cross section speaks in favor of the importance of the role of multinucleon emission process. 
In the present article we will also discuss the role played by relativistic kinematics. The momenta and energies involved in these neutrino reactions are rather large. 
For the MiniBooNE experiment the neutrino energy extends to $\simeq 2$ GeV and the ejected nucleon kinetic energy in a quasielastic process 
can be a few hundred MeV making a non relativistic approximation questionable. 
Indeed, it was pointed out that conclusions on the role of the multinucleon process are doubtful within a non relativistic framework \cite{Amaro:2011qb}. 
It is one of the aims of this work to answer these criticisms. 
To improve our description we introduce in the present work relativistic modifications of the nuclear response, as proposed in \cite{Barbaro:1995ez,DePace:1998yx}. 
To single out their influence we keep for all the remainder of the description the same input parameters 
that we used in our previous work, in particular as concerns the description of the two particle-two hole (2p-2h) processes, 
for which we use our parametrization deduced from the work of Alberico \textit{et al.} \cite{Alberico:1983zg} 
on the 2p-2h contribution to the transverse response. 
We remind that we have also some 3p-3h contribution, taken from \cite{Oset:1987re}. All together we denote the sum by np-nh. 
We also keep the same values of the parameters of the particle-hole (p-h) force which governs 
the collective aspect of the nuclear response via the random phase approximation (RPA). 
We will show that while the relativistic treatment improves the description of the double differential cross section it is nearly 
without influence on the integrated quasielastic cross sections. 
Our previous conclusion on the role played by the multinucleon processes in the axial anomaly is not an artifact of the non relativistic treatment of our earlier works. 
Then we give the single differential cross sections,{\it i.e.}, integrated over the muon energy, or the muon angle, and 
the $Q^2$ distribution not only for charged current (CC) but also for neutral current (NC).

\section{Analysis of differential cross sections}

For a given ``quasielastic'' event the muon energy $E_\mu$ (or kinetic energy $T_\mu$) and its emission angle
$\theta$ are measured. The neutrino energy $E_\nu$ is unknown. In the experimental analysis a specific assumption is made concerning the quasielastic character of the one muon events.
Nuclear cross sections are naturally expressed in terms of the nuclear responses, functions
 of the energy and momentum transferred to the nuclear system, $\omega = E_\nu -E_\mu $, and  $q=|\vec{q}|=|\vec{p_\nu}-\vec{p_\mu}|$. These are the natural variables but they are not the measured quantities. For each value of  $E_\mu$ and $\theta$
 several values of $\omega$, hence of $E_\nu= E_\mu +\omega$, 
 are possible. The expression of the double differential cross section in terms of the measured quantity is 
 
 \begin{equation}
 \label{cross}
\frac{d^2 \sigma}{dT_{\mu}~d~\mathrm{cos}\theta}=
\frac{1}
{ \int \Phi(E_{\nu})~d E_{\nu}}
 \int ~d E_{\nu}
\left[\frac{d^2 \sigma}{d \omega  ~d\mathrm{cos}\theta}\right]_{\omega=E_{\nu}-E_{\mu}} \Phi(E_{\nu}).
\end{equation}
In the numerical evaluations we use the neutrino flux $\Phi(E_{\nu})$ from Ref.\cite{AguilarArevalo:2010zc}.

\begin{figure}
\begin{center}
  \includegraphics[width=16cm,height=12cm]{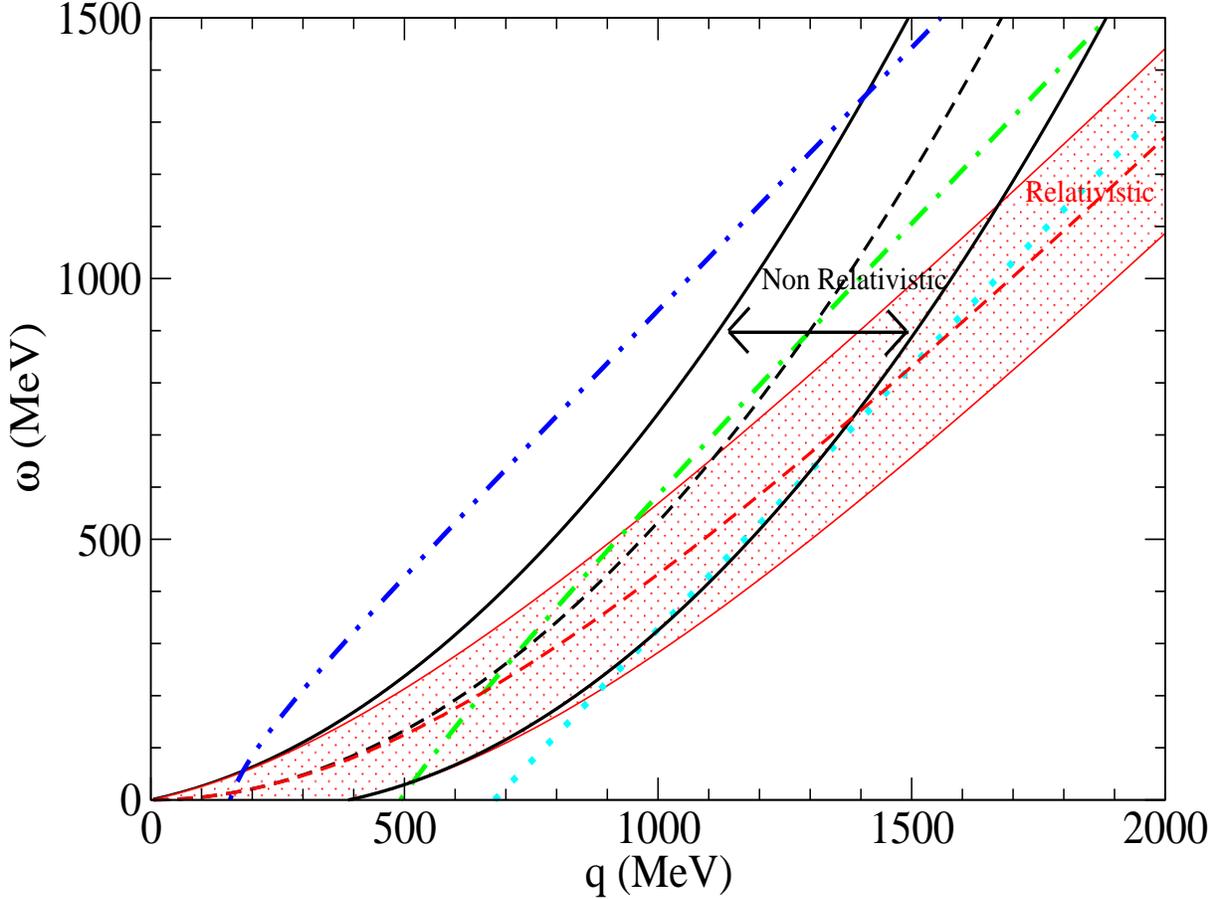}
\caption{(color online). Regions of the quasielastic response of a Fermi gas. 
For relativistic kinematics, see shaded area (red online) delimited by the two corresponding continuous lines. 
In the non relativistic case the horizontal arrow shows the two limiting lines (black online). 
The central dashed lines show the position of the quasielastic peak in the two cases. 
The remnant three lines represent the neutrino 
hyperbolas defined by Eq. (\ref{eq_hyp}) for a muon kinetic energy $T_\mu$=250 MeV and three muon emission angles: cos$\theta$=0.9 (blue dot dot-dashed line), 
cos$\theta$=0 (green dot-dashed line) and cos$\theta$=-0.9 (turquoise dotted line).}
\label{fig_hyperbolas}
\end{center}
\end{figure}

The cross section of the r.h.s. of Eq.(\ref{cross}), as expressed in terms of the nuclear responses \cite{Martini:2009uj}, 
is non vanishing in the regions of the $\omega$  and $q$ plane where the responses are non-zero, 
regions  defined in the following. For a very dilute  Fermi gas the region of response is restricted to the line of the quasi elastic peak, 
namely  $\omega=q^2/(2M_N)$ in the non relativistic case and  $\omega=Q^2/(2M_N)$ for the relativistic kinematics. 
At finite density there is a spreading caused by the Fermi motion and the region of response is delimited, when $q>2k_F$, 
by the two lines $\omega_{\pm}=(q^2 \pm 2qk_F)/(2M_N)$, for the non relativistic case and 
by  $\omega_{\pm}=\sqrt{q^2 \pm 2qk_F +M^2_N} - M_N$,  for the relativistic one. For  $q<2k_F$,  the lower bound is the $\omega =0$ axis in both cases. The two lines delimiting the regions of response are represented in Fig. \ref{fig_hyperbolas}, together with the central one which shows the position of the quasielastic peak, where the response has its maximum, both in the relativistic and non relativistic cases.
 
To illustrate how these regions are explored in neutrino reactions we write the squared four momentum transfer 
in terms of the lepton observables
\begin{equation}
\label{eq_hyp}
Q^2=q^2 - \omega^2= 4 (E_\mu+\omega) E_\mu ~\mathrm{sin} ^2 \frac{\theta}{2}-m_{\mu}^2+2 (E_\mu+\omega) (E_\mu-p_\mu) ~\mathrm{cos} {\theta},
\end{equation} 
with $p_\mu=|\vec{p_\mu}|$. 
For a given set of observables $E_\mu$ and $\theta$ this relation defines a hyperbola  in the  $\omega$ and $q$ plane \cite{Delorme:1985ps}. The asymptotes are parallel to the $\omega=q$ line and the intercept of the curves 
with the $\omega=0$ axis occurs at a value of the momentum
\begin{equation}
q^2_{\textrm{int}}= 4 E_\mu^2 ~\mathrm{sin} ^2 \frac{\theta}{2}-m_{\mu}^2+2 E_\mu (E_\mu-p_\mu) ~\mathrm{cos} {\theta}
\simeq  4 E_\mu^2 ~\mathrm{sin} ^2 \frac{\theta}{2},
\end{equation}
where the second expression is obtained by neglecting the muon mass. 
With increasing $E_\mu$ or increasing  angle, this point shifts away from the origin.
The neutrino cross section for a given $T_\mu$ and $\theta$ explores the nuclear responses along the corresponding hyperbola.
In Fig. \ref{fig_hyperbolas} the quasielastic peak lines are shown together with some examples of hyperbolas.  
This figure illustrates the problems associated with the non relativistic kinematics: the intercept of the hyperbolas with the quasielastic line disappears at large angles, which does not occur in the relativistic case. There can also be two intercepts, which is not realistic either. In order to suppress the pathologies of the non relativistic dynamics and to implement the relativistic corrections 
we use  results from quasielastic electron scattering studies \cite{Barbaro:1995ez,DePace:1998yx}. 
They showed that a good approximation to simulate a relativistic 
treatment starting from a non relativistic frame is obtained with the substitution 
$\omega \to \omega \left(1+\frac{\omega}{2 M_N}\right)$ in the nuclear responses 
(which insures the right position of the quasielastic peak),
and by multiplying the responses by $\left(1+\frac{\omega}{M_N}\right)$. Our present evaluations use these recipes 
and unless specified otherwise the curves of this article are calculated in this framework. Now in a realistic approach of the nuclear dynamics with correlations the nuclear region of response  is not restricted to the Fermi motion band around the quasielastic line (as in Fig. \ref{fig_hyperbolas}) but it covers the whole $\omega$ and $q$ plane due to multinucleon emission. As a consequence, for a given set of values of $E_\mu$ and $\theta$, all values of the energy transfer $\omega$,
hence of the neutrino energy, $E_\nu=E_\mu+\omega$, contribute and one explores the full energy spectrum of neutrinos above the muon energy.

\begin{figure}
\begin{center}
  \includegraphics[width=16cm,height=12cm]{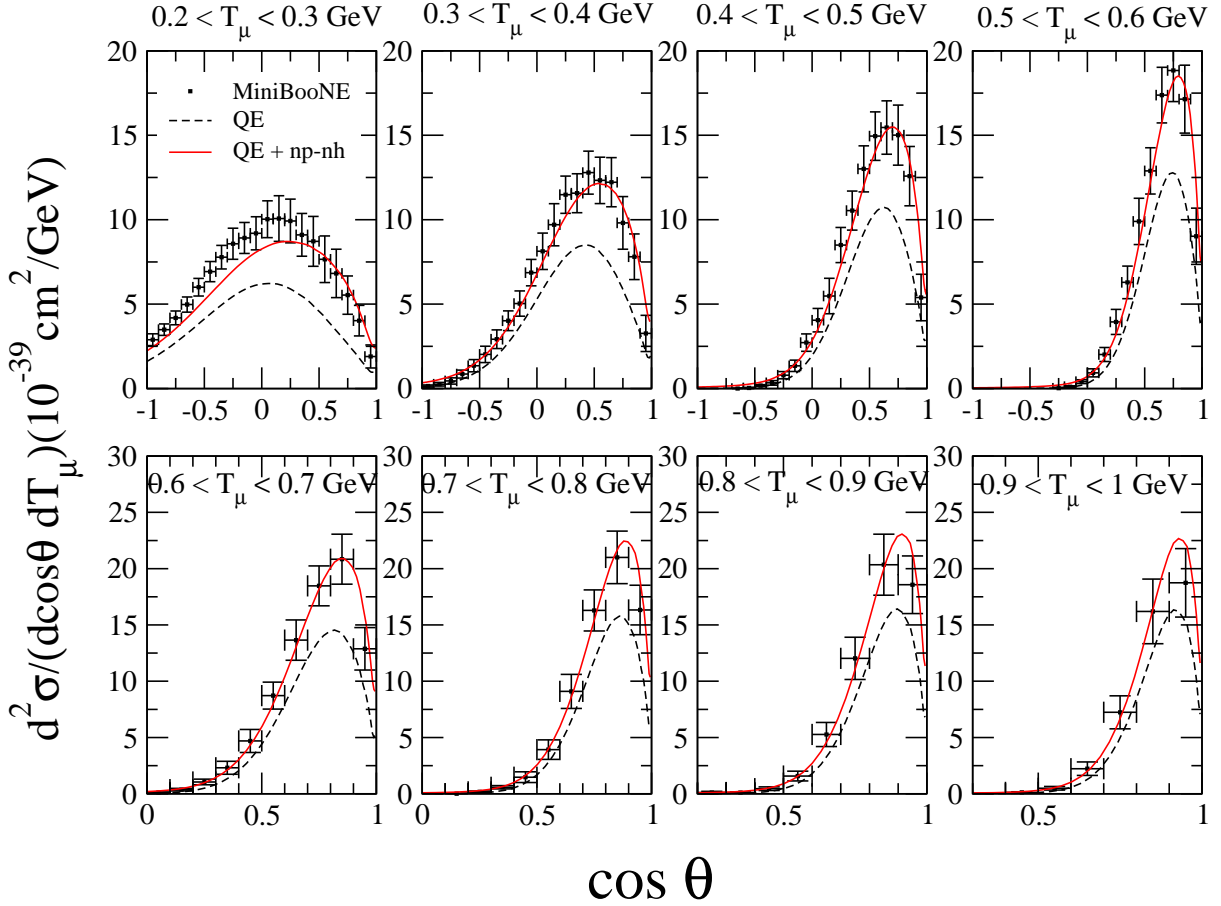}
\caption{(color online). MiniBooNE flux-averaged CC ``quasielastic'' $\nu_\mu$-$^{12}$C double differential cross section per neutron for several values of muon kinetic energy
as a function of the scattering angle. Dashed curve: pure quasielastic (1p-1h) cross section calculated in RPA; 
solid curve: with the inclusion of np-nh component. 
The experimental MiniBooNE points 
are taken from \cite{AguilarArevalo:2010zc}.}
\label{fig_minib_d2s}
\end{center}
\end{figure}

\begin{figure}
\begin{center}
\includegraphics[width=12cm,height=8cm]{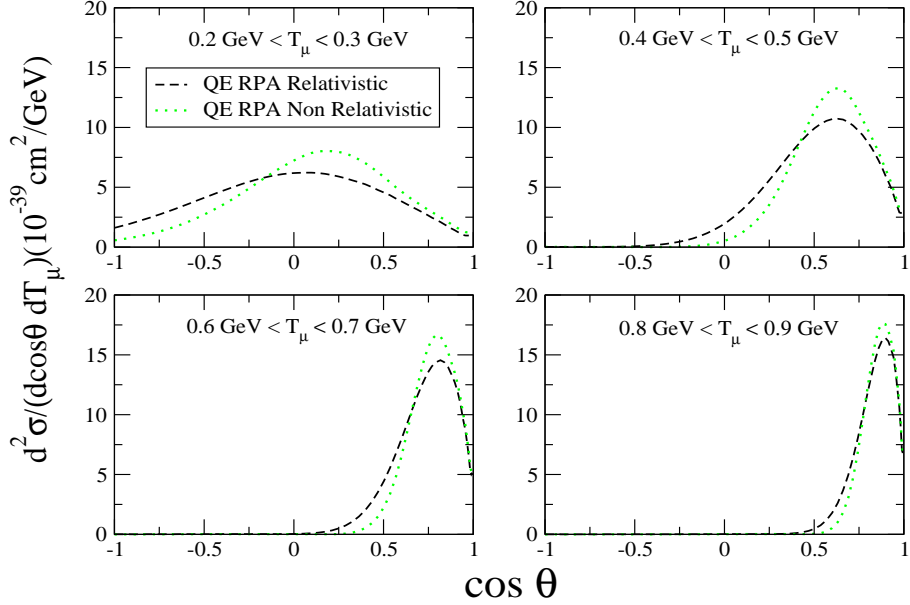}
\caption{(color online). MiniBooNE flux-averaged CC genuine quasielastic $\nu_\mu$-$^{12}$C double differential cross section per neutron for several values of muon kinetic energy
as a function of the scattering angle and calculated in RPA. Dashed curve:  with relativistic corrections; 
dotted curve:  without relativistic corrections.}
\label{fig_minib_d2s_rel_nr}
\end{center}
\end{figure}

\begin{figure}
\begin{center}
\includegraphics[width=12cm,height=8cm]{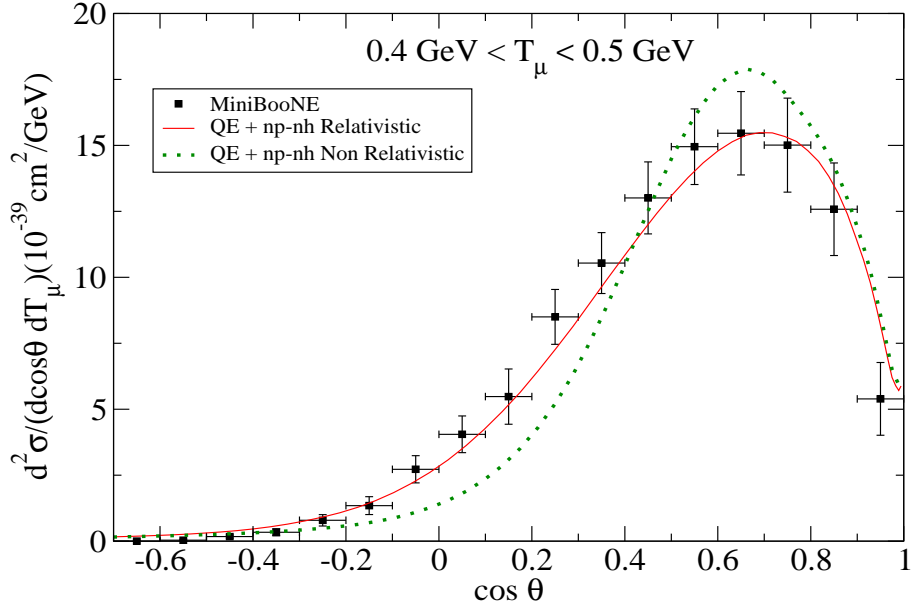}
\caption{(color online). MiniBooNE flux-averaged CC ``quasielastic'' $\nu_\mu$-$^{12}$C double differential cross section 
per neutron for $0.4$ GeV $< T_\mu<0.5$ GeV
as a function of the scattering angle calculated in RPA with the inclusion of the np-nh component. 
Solid curve:  with relativistic corrections; 
dotted curve:  without relativistic corrections.}
\label{fig_minib_d2s_rel_nr_2p2h}
\end{center}

\end{figure}

The results of our present evaluation with the relativistic corrections of  the double differential cross section are displayed 
in Fig. \ref{fig_minib_d2s}, 
with and without the inclusion of the np-nh component and compared to the experimental data. 
This evaluation, as all those of this article, is done with the free value of the axial mass. 
The agreement is quite good in all the measured range once the multinucleon component is incorporated. 
Similar conclusions have been recently reported in \cite{Nieves:2011yp}.
The relativistic corrections are significant, as illustrated in Fig. \ref{fig_minib_d2s_rel_nr} which compares 
the two approaches for the genuine quasielastic contributions. 
The relativistic treatment, which suppresses the kinematical pathologies,  improves the description, in particular in the backward direction. 
This is illustrated in Fig. \ref{fig_minib_d2s_rel_nr_2p2h} in the case $0.4$ GeV $< T_\mu<0.5$ GeV in which the 2p-2h component was added for comparison with data. 
The good agreement with data of Fig. \ref{fig_minib_d2s} is  absent in the non relativistic case. 

\begin{figure}
\begin{center}
  \includegraphics[width=12cm,height=8cm]{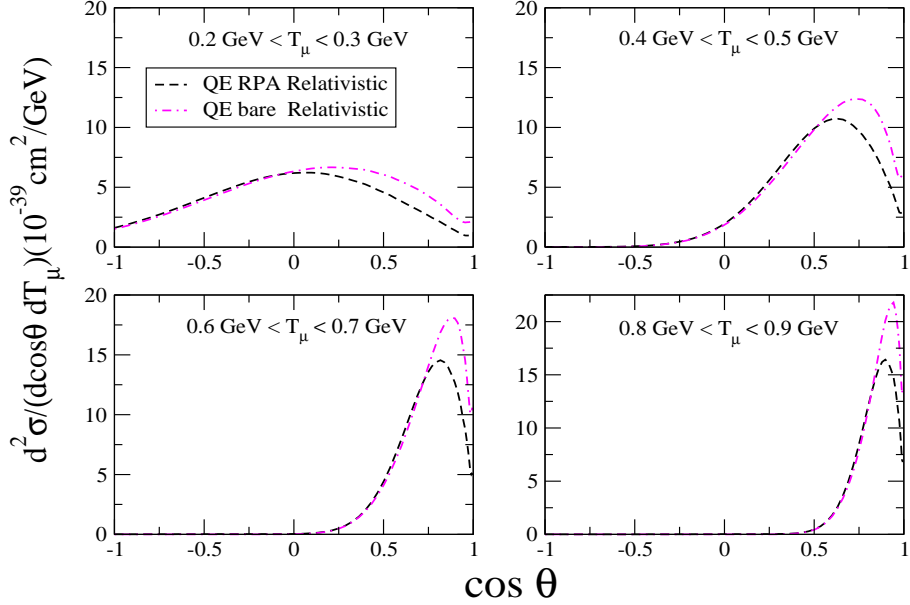}
\caption{(color online). MiniBooNE flux-averaged CC genuine quasielastic $\nu_\mu$-$^{12}$C 
double differential cross section per neutron for several values of muon kinetic energy
as a function of the scattering angle. Dashed curve: calculated in RPA; 
dot-dashed curve: bare.} 
\label{fig_minib_d2s_rpa_bare}
\end{center}
\end{figure}

\begin{figure}
\begin{center}
  \includegraphics[width=12cm,height=8cm]{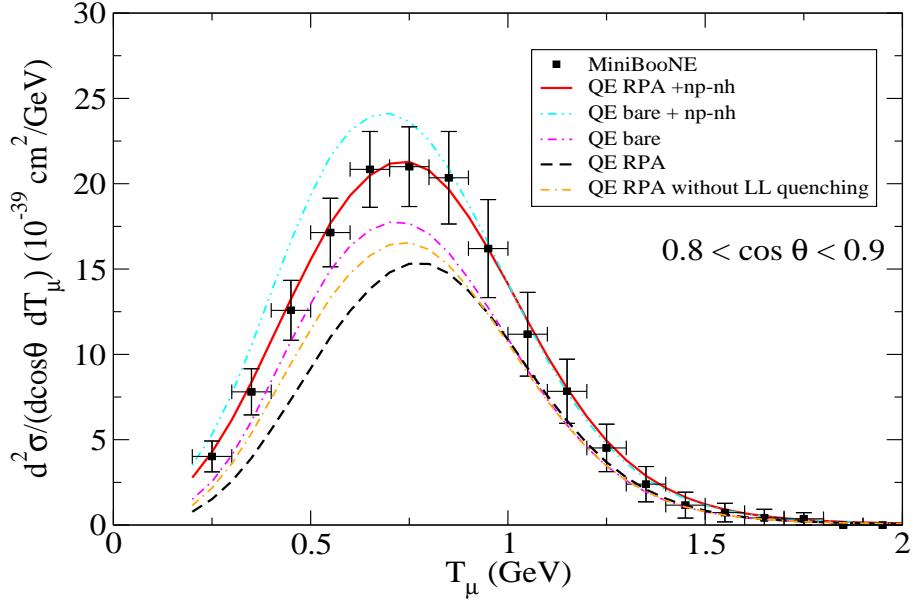}
\caption{(color online). MiniBooNE flux-averaged CC quasielastic $\nu_\mu$-$^{12}$C 
double differential cross section per neutron for 0.8 $<$ cos$\theta~<$ 0.9
as a function of the muon kinetic energy. 
Dashed curve: pure quasielastic calculated in RPA; solid curve: RPA quasielastic 
with the inclusion of np-nh component; dot-dot-dashed: bare quasielastic with the inclusion of np-nh component; 
dot-dashed curve: bare quasielastic; dot-dashed-dashed: RPA quasielastic without the Lorentz-Lorenz (LL) quenching.} 
\label{fig_minib_d2s_cos_fix_vs_tmu}
\end{center}
\end{figure}

\begin{figure}
\begin{center}
\includegraphics[width=5cm,height=6cm]{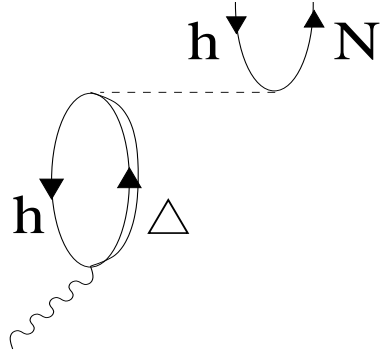}
\caption{Lowest order contribution of Lorentz-Lorenz effect on the quasielastic response.
The wiggled lines represent the external probe, 
and the dashed lines an effective interaction between 
nucleon-hole and $\Delta$-hole.} 
\label{N_Delta_lorentz}
\end{center}
\end{figure}

Our responses  are described, as in our previous works  \cite{Martini:2009uj,Martini:2010ex}, in the framework of random phase approximation. 
Its role is shown in Fig. \ref{fig_minib_d2s_rpa_bare} 
and in Fig. \ref{fig_minib_d2s_cos_fix_vs_tmu}
where the double differential cross sections as a function of 
cos$\theta$ or $T_\mu$ are displayed with and  without RPA.  
The RPA produces a quenching and some shift toward larger angles 
or larger $T_\mu$. 
In Fig. \ref{fig_minib_d2s_cos_fix_vs_tmu} we present the comparison with data adding the np-nh to the genuine QE with or without RPA. 
The fit is significantly better in the RPA framework, reflecting the collective character of the nuclear response. 
The RPA quenching of the cross-sections results from the repulsive nature of the p-h force, embodied in the Landau-Migdal parameter $g'$.
A large part of this quenching arises from the mixing of the p-h states with $\Delta$-hole ones. This
is the Lorentz-Lorenz effect, which concerns exclusively the spin isospin response, hence the axial or magnetic matrix elements. 
In the graphical illustration of the response, the Lorentz-Lorenz effect on the quasielastic one is illustrated in Fig. \ref{N_Delta_lorentz}. 
Figure \ref{fig_minib_d2s_cos_fix_vs_tmu} shows the dominance of the Lorentz-Lorenz quenching effect in the RPA quenching of the neutrino cross-section.  
The Lorentz-Lorenz effect was first predicted for the axial $\beta$ decay matrix elements \cite{Ericson:1973vj}, 
in this case its  existence was controversial. A similar concept was shown by Alberico \textit{et al.} \cite{Alberico:1981sz} 
to apply to the spin-isospin nuclear responses in the region of the quasi elastic peak 
at finite momenta. With the introduction in \cite{Alberico:1981sz} of the collective character 
of the nuclear responses via an RPA treatment, the Lorentz-Lorenz mixing effect naturally appeared. 
However on the experimental side the transverse (magnetic) part of the inclusive electron scattering 
data have not clearly  established the collective nature of this response nor the Lorentz-Lorenz quenching, 
because they are mixed with other effects. 
It is interesting that they seem to show up in neutrino reactions. 
It is in fact the integration over the energy transferred to the nuclear system which is contained 
in these cross sections which allows more easily the emergence of their gross features, an unexpected outcome of these data. 

\begin{figure}
\begin{center}
  \includegraphics[width=12cm,height=8cm]{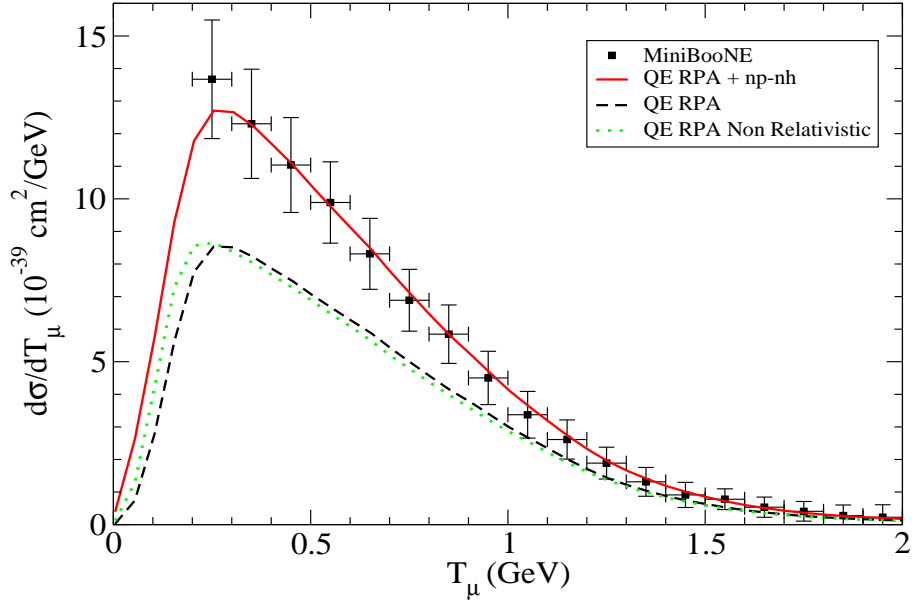}
\caption{(color online). MiniBooNE flux-averaged CC ``quasielastic'' $\nu_\mu$-$^{12}$C differential cross section per neutron 
as a function of the muon kinetic energy. Dashed curve: pure quasielastic (1p-1h) cross section; 
solid curve: with the inclusion of np-nh component; dotted line: pure quasielastic with non-relativistic kinematics. 
The experimental MiniBooNE points 
are taken from \cite{AguilarArevalo:2010zc}.}
\label{fig_minib_ds_dt}
\end{center}
\end{figure}

\begin{figure}
\begin{center}
  \includegraphics[width=12cm,height=8cm]{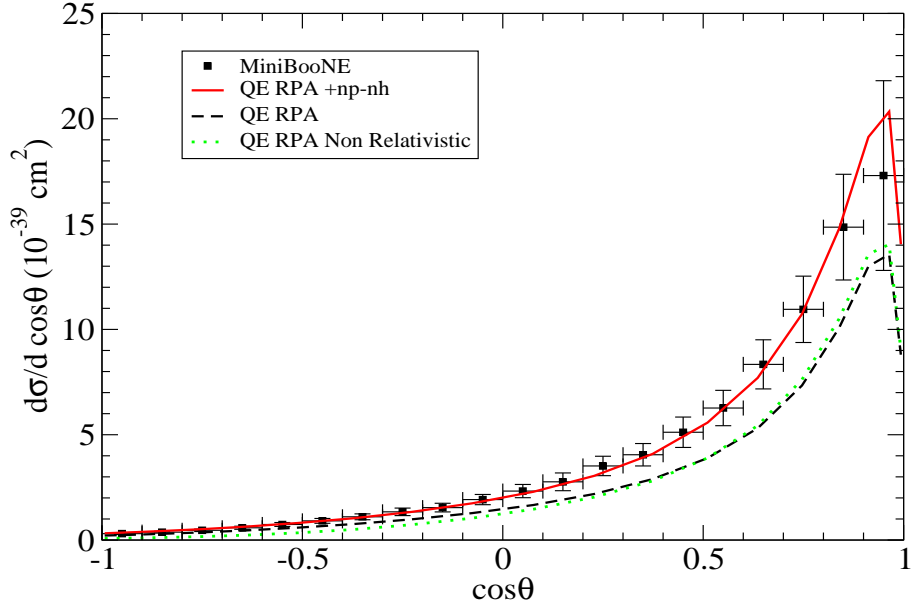}
\caption{(color online). MiniBooNE flux-averaged CC ``quasielastic'' $\nu_\mu$-$^{12}$C differential cross section per neutron 
as a function of the muon scattering angle. 
Note that in order to compare with data the integration is performed over the muon kinetic energies 0.2 GeV $<T_\mu<$  2.0 GeV.
Dashed curve: pure quasielastic (1p-1h) cross section; 
solid curve: with the inclusion of np-nh component; dotted line: pure quasielastic with non-relativistic kinematics. 
The experimental MiniBooNE points 
are taken from \cite{AguilarArevalo:2010zc}.}
\label{fig_minib_ds_dcos}
\end{center}
\end{figure}

In the following we analyze 
 the single differential cross sections, integrated over one of the two independent variables, muon energy or angle, with and without the np-nh contribution. These are displayed in Figs. \ref{fig_minib_ds_dt} and \ref{fig_minib_ds_dcos}.  The agreement of  the integrated cross sections with the data is good in both cases if the multinucleon contribution is included, as displayed in Figs \ref{fig_minib_ds_dt} and \ref{fig_minib_ds_dcos}.
Here, contrary to the previous case,  the relativistic corrections have a small influence.  
A single integration nearly washes out the relativistic effects. This is a fortiori true for the total cross section.

\begin{figure}
\begin{center}
  \includegraphics[width=12cm,height=8cm]{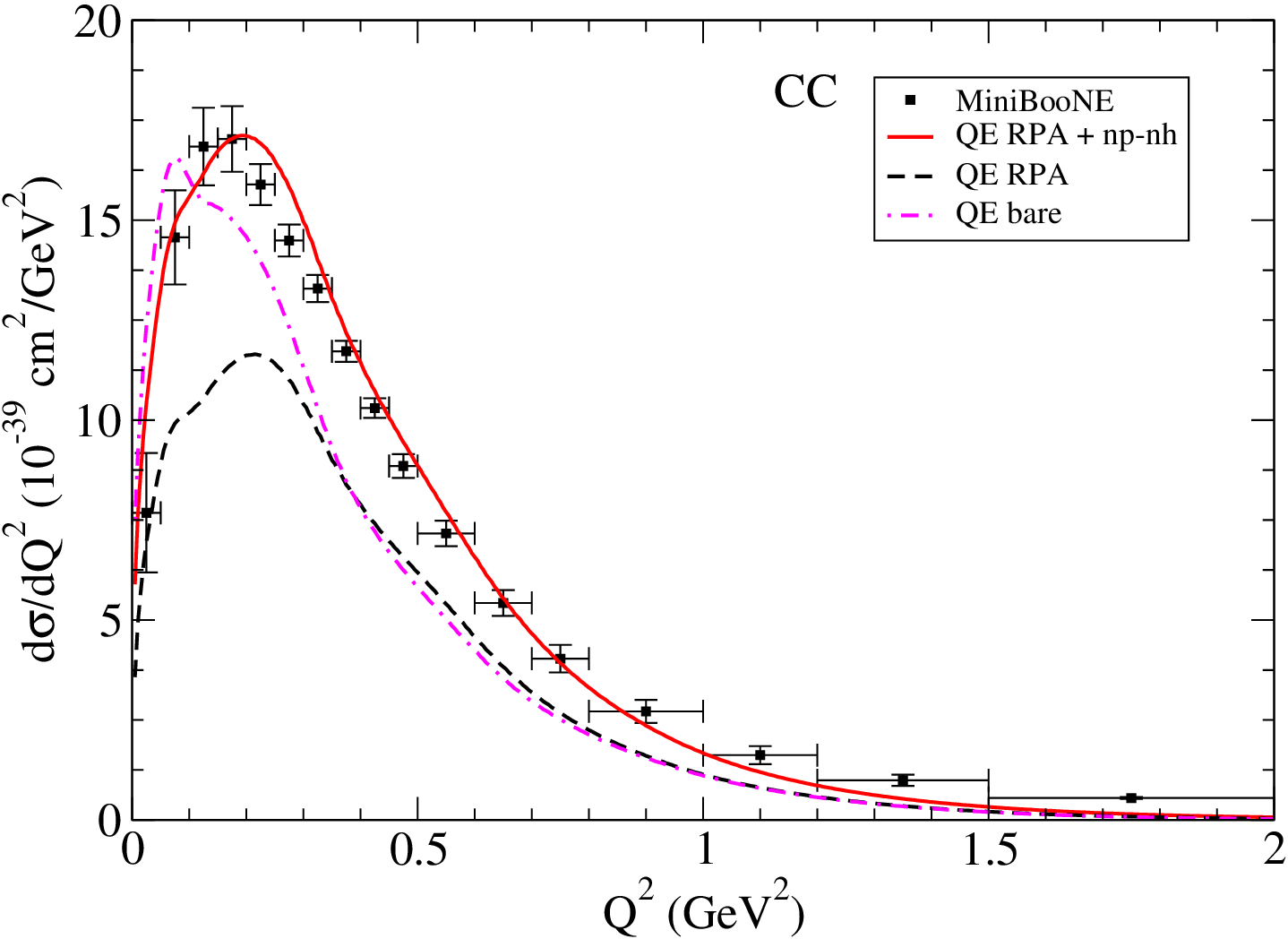}
\caption{(color online). MiniBooNE flux-averaged CC $Q^2$ distribution per neutron. Dashed curve: pure quasielastic (1p-1h); 
solid curve: with the inclusion of np-nh component; dot-dashed line: bare distribution. 
The experimental MiniBooNE points 
are taken from \cite{AguilarArevalo:2010zc}.}
\label{fig_minib_cc_ds_dQ2}
\end{center}
\end{figure}

Finally we show in Fig. \ref{fig_minib_cc_ds_dQ2} the single differential cross section with respect to $Q^2$, 
which was historically of interest for the determination of the axial form factor. 
Figure \ref{fig_minib_cc_ds_dQ2} shows that here also the fit is good without any modifications of the axial mass but provided the multinucleon component is incorporated.
 In this figure the existence of a large  $Q^2$ region, $Q^2 \gtrsim .2$ GeV$^2$, in which RPA is practically without influence is of great interest. It allows to single out the need for the np-nh contribution without any interference from the RPA effects. At low  $Q^2$ where the bare description without np-nh is able to reproduce the data the RPA quenching is then needed to compensate the enhancement from the  
np-nh contribution. We remind that, in  the absence of  a quantitative evidence in the electron scattering data for the collective nature or the Lorentz-Lorenz quenching, 
the parameters that govern this effect are not in full control. The neutrino experiments will be of great help to narrow their range. 

\begin{figure}
\begin{center}
  \includegraphics[width=12cm,height=8cm]{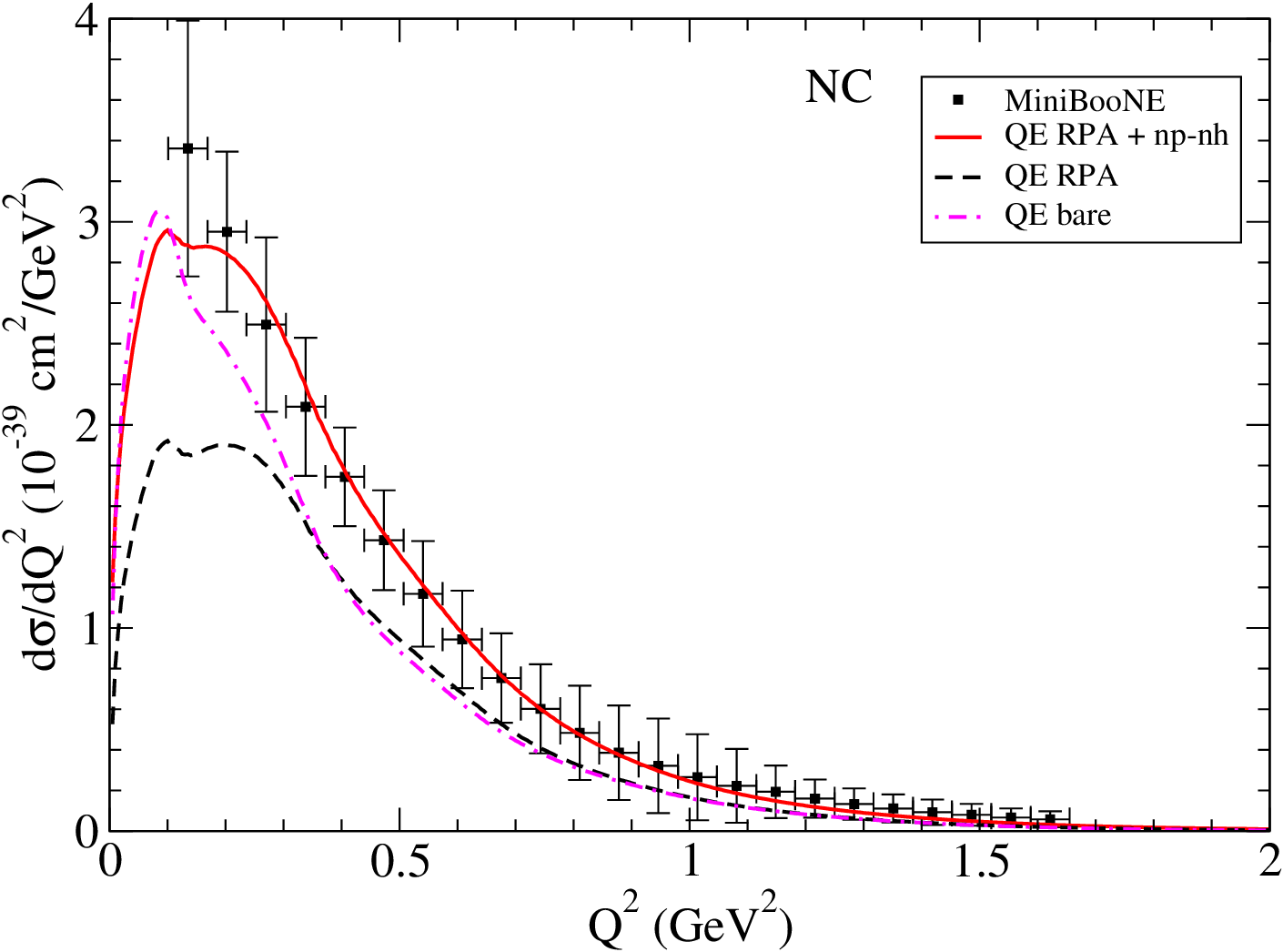}
\caption{(color online). MiniBooNE flux-averaged NC $Q^2$ distribution per nucleon. Dashed curve: pure quasielastic (1p-1h); 
solid curve: with the inclusion of np-nh component; dot-dashed line: bare distribution. 
The experimental MiniBooNE points 
are taken from \cite{AguilarArevalo:2010cx}.}
\label{fig_minib_nc_ds_dQ2}
\end{center}
\end{figure}

On the experimental side the same differential cross section but for neutral currents was recently published \cite{AguilarArevalo:2010cx}. 
It is interesting to compare it 
to our predictions. Here the final lepton, a neutrino, is not observed and the transfer variable, $Q^2$ is obtained indirectly from the kinetic energy of the  ejected nucleons. In this case it is not quite clear how the multinucleon component shows up in the experimental data. However the same problem of the axial mass also seems  to emerge from these data \cite{Benhar:2011wy,Butkevich:2011fu}.
We have thus confronted our theory with the published $Q^2$ distribution. The data are for CH$_2$ instead of pure carbon as in our theory, but 
the difference between the two cases was shown to be small \cite{Benhar:2011wy}. 
The comparison of our evaluation with data is shown in Fig. \ref{fig_minib_nc_ds_dQ2}. 
It turns out that the combination of RPA quenching and 2p-2h piece leads to a good agreement with data.

\section{Conclusions}
In conclusion we have investigated in this work more in detail the neutrino-$^{12}$C cross section in connection with MiniBooNE data. 
The most significant quantity is the double differential cross section which does not imply any reconstruction of the neutrino energy. 
To compare our theoretical model to these data we have improved our original description applying relativistic corrections. 
The agreement of our RPA approach with data is quite good once the np-nh component is included. 
It confirms our first suggestion that there is no need for a change in the axial mass once the multinucleon processes are taken into consideration. 
A good agreement is also found for the (single) differential cross sections integrated over one variable where the relativistic corrections play practically no role. 
We have also examined the $Q^2$ distribution which establishes the necessity of the multinucleon contribution, independently of the RPA quenching.
The same description appears to be efficient for the $Q^2$ distribution in the case of neutral currents, 
although  the role of the multinucleon component in these experimental data is not obvious.
Understanding in detail the role of nuclear dynamics in neutrino-nucleus interactions is important for the neutrino oscillation programs (see for example \cite{FernandezMartinez:2010dm}) but  beyond the question of the axial mass which is our main goal, our study also has an interest from a purely hadronic viewpoint. The fact that a signature for the RPA influence in the form of the Lorentz-Lorenz quenching, a long-sought-after effect, seems to emerge from neutrino reactions is an additional and unexpected outcome of our study.



\begin{thebibliography}{99}
\bibitem{AguilarArevalo:2010zc}
  A.~A.~Aguilar-Arevalo {\it et al.} [ MiniBooNE Collaboration ],
  Phys.\ Rev.\  {\bf D81}, 092005 (2010).

\bibitem{AguilarArevalo:2010cx}
  A.~A.~Aguilar-Arevalo {\it et al.} [ MiniBooNE Collaboration ],
  Phys.\ Rev.\  {\bf D82}, 092005 (2010).


\bibitem{Martini:2009uj}
  M.~Martini, M.~Ericson, G.~Chanfray, J.~Marteau,
  Phys.\ Rev.\  {\bf C80}, 065501 (2009).

\bibitem{Martini:2010ex}
  M.~Martini, M.~Ericson, G.~Chanfray, J.~Marteau,
  Phys.\ Rev.\  {\bf C81}, 045502 (2010).

\bibitem{Benhar:2010nx}
  O.~Benhar, P.~Coletti, D.~Meloni,
  Phys.\ Rev.\ Lett.\  {\bf 105}, 132301 (2010).

\bibitem{Amaro:2010sd}
  J.~E.~Amaro, M.~B.~Barbaro, J.~A.~Caballero, T.~W.~Donnelly, C.~F.~Williamson,
  Phys.\ Lett.\  {\bf B696}, 151-155 (2011).
  


\bibitem{Amaro:2011qb}
  J.~E.~Amaro, M.~B.~Barbaro, J.~A.~Caballero, T.~W.~Donnelly, J.~M.~Udias,
  Phys.\ Rev.\  {\bf D84}, 033004 (2011).

\bibitem{Nieves:2011pp}
  J.~Nieves, I.~Ruiz Simo, M.~J.~Vicente Vacas,
  Phys.\ Rev.\  {\bf C83}, 045501 (2011).


\bibitem{Nieves:2011yp}
  J.~Nieves, I.~R.~Simo, M.~J.~V.~Vacas,
  [arXiv:1106.5374 [hep-ph]].

\bibitem{Bodek:2011ps}
  A.~Bodek, H.~Budd, M.~E.~Christy, 
  Eur.\ Phys.\ J.\  {\bf C71}, 1726 (2011).


\bibitem{Meucci:2011vd}
  A.~Meucci, M.~B.~Barbaro, J.~A.~Caballero, C.~Giusti, J.~M.~Udias,
Phys.\ Rev.\ Lett.\  {\bf 107}, 172501 (2011).

\bibitem{Barbaro:1995ez}
  M.~B.~Barbaro, A.~De Pace, T.~W.~Donnelly, A.~Molinari,
  Nucl.\ Phys.\  {\bf A596}, 553-585 (1996).

\bibitem{DePace:1998yx}
  A.~De Pace,
  Nucl.\ Phys.\  {\bf A635}, 163 (1998).


\bibitem{Alberico:1983zg}
  W.~M.~Alberico, M.~Ericson, A.~Molinari,
  Annals Phys.\  {\bf 154}, 356 (1984).

\bibitem{Oset:1987re} E.~Oset and L.~L.~Salcedo, Nucl.\ Phys.\  A {\bf 468} 631 (1987).

\bibitem{Delorme:1985ps}
  J.~Delorme, M.~Ericson,
  Phys.\ Lett.\  {\bf B156}, 263 (1985).
  

\bibitem{Ericson:1973vj}
  M.~Ericson, A.~Figureau, C.~Thevenet,
  Phys.\ Lett.\  {\bf B45}, 19-22 (1973).



\bibitem{Alberico:1981sz}
  W.~M.~Alberico, M.~Ericson, A.~Molinari,
  Nucl.\ Phys.\  {\bf A379}, 429 (1982).


\bibitem{Benhar:2011wy}
  O.~Benhar, G.~Veneziano,
  Phys.\ Lett.\  {\bf B702}, 433-437 (2011).

\bibitem{Butkevich:2011fu}
  A.~V.~Butkevich, D.~Perevalov,
  Phys.\ Rev.\  {\bf C84}, 015501 (2011).



\bibitem{FernandezMartinez:2010dm}
  E.~Fernandez-Martinez, D.~Meloni,
  Phys.\ Lett.\  {\bf B697}, 477-481 (2011).



\end{thebibliography}
\end{document}